\documentclass[a4paper, 11pt]{article}
\usepackage{amsmath,amssymb,color,graphics,epsfig,cite}

\usepackage{amsfonts}
\def\d{{\rm d}}

\renewcommand{\log}{\textup{log}}
\renewcommand{\exp}{\textup{e}}

\newcommand{\half}{{\textstyle{\frac{1}{2}}}}

\newcommand{\go}{\overset{\circ}{g}}
\newcommand{\Go}{\overset{\circ}{G}}
\newcommand{\Wo}{\overset{\circ}{W}}
\newcommand{\So}{\overset{\circ}{S}}
\newcommand{\Eo}{\overset{\circ}{E}}
\newcommand{\omo}{\overset{\circ}{\omega}}
\newcommand{\Gao}{\overset{\circ}{\Gamma}}
\newcommand{\zeo}{\overset{\circ}{\zeta}}

\makeatletter
\@addtoreset{equation}{section}
\makeatother

\begin{document}

\begin{titlepage}
\hfill 
\vspace{2.5cm}
\begin{center}

{{\LARGE  \bf Entanglement entropy, the Einstein equation and the
Sparling construction}} \\

\vskip 1.5cm {\large Mahdi Godazgar}
\\
{\vskip 0.5cm
\textit{ Institut f\"ur Theoretische Physik,\\
Eidgen\"ossische Technische Hochschule Z\"urich, \\
Wolfgang-Pauli-Strasse 27, 8093 Z\"urich, Switzerland.}\\[5mm]
godazgar@phys.ethz.ch}
\end{center}

\vskip 0.35cm

\begin{center}
\today
\end{center}

\noindent

\vskip 1.2cm

\begin{abstract}
\noindent We relate the recent derivation of the linearised Einstein equation on an AdS background from holographic entanglement entropy arguments to the Sparling construction: we derive the differential form whose exterior derivative gives the Einstein equation from the Sparling formalism.  We develop the study of perturbations within the context of the Sparling formalism and find that the Sparling form vanishes for linearised perturbations on flat space.
\end{abstract}

\end{titlepage}

\section{Introduction}
One of the main puzzles of AdS/CFT, or holography in general, is how the bulk geometry and, in particular, bulk locality arises from data in the CFT in question.  This has lead to much interest recently.  One interesting string of ideas is whether considering entanglement between different regions on the boundary theory may be used to discern important properties of the bulk geometry.~\footnote{For example, see Ref.~\cite{holent} and references therein.}  

In the context of general relativity, a seemingly unrelated problem is the suitable definition of energy.  The equivalence principle precludes a meaningful local definition of energy.  However, one would hope to be able to define a vector that measures the energy-momentum in some region enclosed by, say, a ball of radius $r>0$---a quasi-local definition of energy.  Beyond the scope of stationary asymptotically flat/AdS spaces, it is not even clear how to formulate a suitable definition of energy for a whole space.~\footnote{For example, see Ref.~\cite{energyGR} and references therein.}

The idea behind this paper is that these two important problems may, in fact, be related.

In recent work \cite{ent1, ent2, ent3}, the authors argue that the linearised Einstein equations on a $(d+1)$-dimensional anti-de Sitter background can be derived by considering the change in the entanglement entropy for a ball-shaped region $A$ under a perturbation of the vacuum state of the boundary CFT in a holographic set-up.  The starting point in the derivation is the first law of entanglement entropy \cite{ent2}, which states that the change in the entanglement entropy associated to region $A$ is equal to the change in the expectation value of the modular, or entanglement, Hamiltonian.  This first law is then translated to a gravitational first law that applies to an AdS-Rindler horizon in the bulk constructed using the Ryu-Takayanagi \cite{Ryu} prescription.  This gravitational first law can be thought of as the AdS-Rindler analogue of the Iyer-Wald first law for asymptotically flat black hole horizons \cite{iyer}.  Then, the challenge of deriving the Einstein equation essentially translates to reversing the Iyer-Wald theorem that gives the gravitational first law from the Einstein equation \cite{ent1, ent3}.  The result is that the linearised Einstein equation is given as the exterior derivative of a $(d-1)$-form.  The fact that the Einstein equation can be written in terms of the exterior derivative of a differential form is characteristic of the Sparling construction \cite{sparling1, sparling2, mason, dubois}.

The Sparling construction can be best thought of, at least for our purposes, as a construction in the orthonormal frame bundle that gives the Einstein tensor as the sum of the exterior derivative of a $(d-2)$-form, the Witten-Nester form and a $(d-1)$-form, the Sparling form.  Importantly, the two differential forms that enter this equation are not horizontal.  Hence, the construction depends non-covariantly on the choice of section.  An important corollary of this construction is that given an orthonormal frame such that the associated metric satisfies the vacuum Einstein equation, the Sparling form becomes exact and equal to the exterior derivative of the Witten-Nester form.  Both of these pseudo-tensors play an important r\^ole in the understanding of energy in general relativity, such as the Witten proof of the positive ADM mass theorem and similar positivity of mass proofs, Penrose's definition of quasi-local energy and the Hamiltonian of general relativity as expressed through the Ashtekar variables (see Ref.~\cite{mason}).

In this paper, we relate the recent holographic derivation of the linearised Einstein equation to the Sparling construction by rederiving the differential form whose exterior derivative gives the linearised Einstein equation from the Sparling formalism.  We demonstrate this correspondence explicitly for four dimensions only in the interests of clarity.  A similar analysis for higher dimensions will be straightforward and along essentially the same lines as that used in four dimensions.  An important ingredient in demonstrating this relation is formulating linearised perturbation theory within the Sparling formalism.  Rather interestingly, though not surprisingly, we find that for linearised perturbations around flat space the Sparling form vanishes identically and the linearised Einstein equation is always given as the exterior derivative of the perturbed Witten-Nester form.

In section \ref{sec:ent}, following Ref.~\cite{ent3} we review the derivation of the linearised Einstein equation from holographic entropy arguments \cite{ent1, ent3} and in section \ref{sec:sparling}, we review the Sparling construction.  In section \ref{sec:pertspa}, we develop the formalism needed to consider perturbations in the context of the Sparling construction and briefly consider perturbations on flat space in section \ref{sec:flat}, before moving on to the main interest of the paper in section \ref{sec:ads}: perturbations on AdS space, where we also derive the linearised Einstein equation as the exterior derivative of a two-form and show that this is the same two-form as that found in Ref.~\cite{ent3} up to an exact form.  We end with some conclusions and outlook.

\section{Holographic entanglement and the Einstein equation} \label{sec:ent}

In this section, following Ref.\ \cite{ent3}, we briefly review the recent derivation of the linearised Einstein equation using holographic entanglement entropy ideas \cite{ent1, ent3}.  The philosophy in this construction is to address the bulk locality puzzle by arguing that the linearised Einstein equation around an anti-de Sitter background follows from small perturbations of the CFT vacuum state.  Although, it turns out that the actual derivation of the linearised equation is independent of holography, holography justifies the gravitational ``first law'' that is used to derive the Einstein equation.

The starting point is the first law of entanglement entropy \cite{ent2}
\begin{equation} \label{1law}
 \delta S_A = \delta \langle H_A \rangle ,
\end{equation}
where $\delta S_A$ is the first-order change in the entanglement entropy for a region $A$, while the right hand side is the first-order variation in the expectation value of the modular, or entanglement, Hamiltonian $H_A$.  Both the entanglement entropy and Hamiltonian are defined via the reduced density matrix associated with region $A$
\begin{equation}
 \rho_A = tr_{\bar{A}} (\rho);
\end{equation}
\begin{equation}
 S_A = - tr (\rho_A \, \log \rho_A), \qquad \rho_A = \frac{\exp^{-H_A}}{tr(\exp^{-H_A})}.
\end{equation}
It is not too difficult to derive the first law \eqref{1law} from the definitions above as well as the constraint that the reduced density matrix is unit trace \cite{ent3}.  

Now that we have such a law, which resembles/generalises the first law of thermodynamics, an obvious question in the context of holography is how this translates to a gravitational statement in the bulk.  More precisely, we assume that the gravitational state corresponding to the CFT vacuum state is $(d+1)$-dimensional anti-de Sitter space in Poincar\'{e} coordinates
\begin{equation} \label{met:ads}
\d s^2 =g_{\mu \nu} \d x^\mu \d x^\nu = \frac{\ell^2}{z^2} \left( \d z^2 + \eta_{\alpha \beta}\, \d X^\alpha \d X^\beta \right)
\end{equation}
with index $\mu=(z,\alpha)$ and $X^\alpha = (t, x^{\hat{\alpha}})$.  A perturbation of the CFT vacuum state is going to correspond, holographically, to a perturbed geometry about the AdS background.  The question, then, is what does the first law above, which constrains the admissible perturbations on the boundary, imply for the gravitational perturbations in the bulk?

In general, this turns out to be a difficult problem.  However, the case where the region $A=B(R,x_0)$, corresponding to a ball of radius $R$, centre $x_0$, is well-understood \cite{Hislop:1981uh,CHM11}.  Here, one identifies $S_A$ with $S^{grav}$, the entropy associated with an AdS-Rindler wedge $\tilde{B}$ at ``temperature'' $T=1/(2\pi R)$, such that $\partial \tilde{B} = \partial B.$  Note that the boundary surfaces $B$ and $\tilde{B}$ enclose a constant $t$ hypersurface $\Sigma$.  Moreover, the gravitational analogue of $\langle H_A \rangle$ is the canonical energy associated with the Killing vector $\xi$ that generates the Rindler horizon, which we denote $E[\xi]$.  On the hypersurface $\Sigma$, $\xi \propto \partial_t$.  Thus, in conclusion, we have a gravitational statement to the effect that
\begin{equation}
 \delta S^{grav} = \delta E[\xi].
\end{equation}
Were we considering an asymptotically flat stationary black hole solution with a bifurcate Killing horizon, \textit{i.e.} non-zero surface gravity $\kappa$, normalised to $\kappa = 2\pi$, and a static solution of the linearised Einstein equation around the black hole background, then the above identity is the content of the Iyer-Wald theorem \cite{iyer}.  Therefore, essentially, what we are hoping to achieve is the reverse of the Iyer-Wald theorem applied to an AdS-Rindler background.

A clue as to how to proceed is that the entropy, whether in the context of Einstein gravity where it corresponds to the area of the horizon or more generally, where it is given by the Wald prescription, is given by an integral over the horizon, in this case $\tilde{B}$.  Similarly, the canonical energy, as is to be expected of energy definitions in gravity, is given by an integral over the boundary of the space; in this case anti-de Sitter space.  As long as it is independent of the surface of integration, then we may define it as an integral over the surface $B.$  If the integrands in the two integrals are the same then we can use Stokes' theorem to relate their difference to an integral on $\Sigma.$

The Iyer-Wald formalism provides a $(d-1)$-form $\chi$, the integrand of the presymplectic form, such that
\begin{equation} \label{SEchi}
 \delta S^{grav} = \frac{1}{16 \pi G_{N}} \int_{\tilde{B}} \chi, \qquad \delta E[\xi] = \frac{1}{16 \pi G_{N}} \int_{B} \chi
\end{equation}
and
\begin{equation} \label{linEinstein}
 \d \chi = - 2 \xi^\mu \delta G_{\mu \nu} \epsilon^\nu,
\end{equation}
where $\delta G_{\mu \nu}$ is the linearised Einstein equation and $\epsilon^\nu$ is the volume form on a surface with normal vector $\partial/\partial X^{\nu}.$  

Moreover, the conservation and tracelessness of the CFT stress tensor gives that 
\begin{equation}
 \d \chi = 0
\end{equation}
on the AdS boundary, corresponding to the surface $z=0,$ so that $\delta E[\xi]$ is independent of the surface of integration.  The above ingredients imply that
\begin{align}
 0 &= \delta S^{grav} - \delta E[\xi] \notag \\
   &= \frac{1}{16 \pi G_{N}}\int_{\tilde{B}} \chi - \frac{1}{16 \pi G_{N}} \int_{B} \chi \notag \\
   &= \frac{1}{16 \pi G_{N}} \int_{\Sigma} \d \chi \notag \\
   &= - \frac{1}{8 \pi G_{N}} \int_{\Sigma} \xi^t \delta G_{tt} \epsilon^t,
\end{align}
where in the last line we have only the $t$-components of $\xi^{\mu}$ and $\epsilon^\nu$ contributing, because these are the only non-zero components on $\Sigma.$  Since, $\Sigma$ is arbitrary, we conclude that
\begin{equation}
 \delta G_{tt} = 0.
\end{equation} 
The above result was derived by considering a ball $B$ in a constant $t$ slice.  However, we can equivalently consider another frame of reference and the above argument will go through all the same.  Thus,
\begin{equation}
 \delta G_{\alpha \beta} = 0.
\end{equation}
The remaining components of the linearised Einstein equation are constraint equations in a radial slicing of the space formulated as an initial value problem.  Thus, as long as they are satisfied on the $z=0$ surface, which they can be shown to be \cite{ent3}, then they hold for all values of $z$.  This completes the derivation of the linearised Einstein equation from the first law of entanglement entropy, but most importantly, as far as we are concerned, it relates the linearised Einstein equation to the exterior derivative of some $(d-1)$-form provided by Iyer and Wald, eqn.~\eqref{linEinstein}.

For Einstein gravity \cite{ent3}
\begin{equation}
 \chi = \delta(\nabla^{\mu} \xi^{\nu}) \epsilon_{\mu \nu},
\end{equation}
where $h_{\mu\nu}=\delta g_{\mu\nu}$, the traceless and transverse perturbed metric, is defined via
\begin{equation}
 g_{\mu\nu} = \go_{\mu\nu} + h_{\mu\nu}
\end{equation}
with background metric\footnote{In general, we denote all background quantities with a circle on top, except when it is clear from the context.  For example, in the expression $\nabla_{\mu} h_{\nu\rho},$ it is clear that the covariant derivative is with respect to the background metric in order for the expression to remain first order.} $\go_{\mu\nu}$ and
\begin{equation} \label{eps2}
 \epsilon_{\mu\nu} = \frac{1}{(d-1)!} \epsilon_{\mu\nu \rho_1 \dots \rho_{(d-1)}} dX^{\rho_1} \wedge \ldots \wedge dX^{\rho_(d-1)}.
\end{equation}
Choosing to work in a radial gauge,
\begin{equation}
 h_{\mu z} = 0
\end{equation}
we find that on $\Sigma=\{ t=t_0\}$ \cite{ent3}
\begin{align} 
 \chi|_{\Sigma} &= - \xi^t \left\{ \partial_z h^{\hat{\alpha}}{}_{\hat{\alpha}}\, \epsilon^{t}{}_z - \left( \partial^{\hat{\alpha}} h^{\hat{\beta}}{}_{\hat{\beta}} -  \partial^{\hat{\beta}} h^{\hat{\alpha}}{}_{\hat{\beta}} \right) \epsilon_{t \hat{\alpha}} \right\} \notag \\[2mm]
  & \qquad - \frac{2\pi}{R} \left\{ z h^{\hat{\alpha}}{}_{\hat{\alpha}}\, \epsilon^{t}{}_z + \left[ (x^{\hat{\alpha}} - x_0^{\hat{\alpha}}) h^{\hat{\beta}}{}_{\hat{\beta}} 
  - (x^{\hat{\beta}} - x_0^{\hat{\beta}}) h^{\hat{\alpha}}{}_{\hat{\beta}} \right] \epsilon^{t}{}_{\hat{\alpha}} \right\},
  \label{chi}
\end{align}
where we have used the fact that in Poincar\'e coordinates
\begin{equation}
 \xi = \frac{\pi}{R} \left\{ \left[ R^2 - z^2 - (t-t_0)^2 - (\underline{x} - \underline{x}_0)^2 \right] \partial_t 
 - 2(t-t_0) \left[ z \partial_z + (x^{\hat{\alpha}} - x_0^{\hat{\alpha}}) \partial_{\hat{\alpha}} \right] \right\}.
\end{equation}
Note that on $\Sigma$, only the $t$-component of $\xi$ is non-zero.

Upon taking the exterior derivative of $\chi$, the second term on the right hand side of eqn.~\eqref{chi} cancels the derivative of $\xi^t$ in the first term, so that
\begin{equation}
 \d \chi |_{\Sigma} = - \xi^t \d \left\{ \partial_z h^{\hat{\alpha}}{}_{\hat{\alpha}}\, \epsilon^{t}{}_z - \left( \partial^{\hat{\alpha}} h^{\hat{\beta}}{}_{\hat{\beta}} -  \partial^{\hat{\beta}} h^{\hat{\alpha}}{}_{\hat{\beta}} \right) \epsilon_{t \hat{\alpha}} \right\}.
\end{equation}
In summary, on $\Sigma$
\begin{equation} \label{ent:Einstein}
  \delta G_{tt}\, \xi^t \epsilon^t = \half \xi^t \d \left\{ \partial_z h^{\hat{\alpha}}{}_{\hat{\alpha}}\, \epsilon^{t}{}_z - \left( \partial^{\hat{\alpha}} h^{\hat{\beta}}{}_{\hat{\beta}} -  \partial^{\hat{\beta}} h^{\hat{\alpha}}{}_{\hat{\beta}} \right) \epsilon_{t \hat{\alpha}} \right\}.
\end{equation}

\section{The Sparling form} \label{sec:sparling}

In general relativity, the equivalence principle means that a local definition of energy is impossible. Given that the equations are second-order, one would expect the energy-momentum density at a point to be first order in the gravitational field.  However, a local coordinate transformation can then be used to set this to zero.  Thus, a reasonable expectation is that a quasi-local definition of energy-momentum ought to be pseudo-tensorial.  As overwhelming as this may seem, one could view the pseudo-tensors in the different frames as being pull-backs in different local sections of some bundle on which a canonical expression for the energy-momentum is defined.  Indeed, this was the motivation for Sparling's construction \cite{sparling1,sparling2,mason}, which we review in this section.\footnote{The original construction of Sparling's is defined on the spin bundle. However, for our purposes it will be more useful to work with the orthonormal frame bundle \cite{F89}.}  Although the original construction is for a four-dimensional space, one can simply construct similar objects in higher dimensions \cite{dubois}.  However, here, we keep to four dimensions, since this is sufficient to get the main ideas across without introducing more notation, albeit simple.

Consider an orthonormal frame $\theta^a$.  The Cartan equations, for vanishing torsion read
\begin{gather}
 \d \theta^a + \omega^{a}{}_{b}\wedge \theta^b=0, \label{cartan1} \\
 \d \omega^a{}_{b} + \omega_a{}_c \wedge \omega^c{}_b = \Omega^a{}_{b},
 \label{cartan2}
\end{gather}
where $\omega_{ab}$ is the spin connection, which we choose to be anti-symmetric.  This corresponds to a choice of a metric compatible connection 
\begin{equation} \label{metcom}
 \partial_{\mu} \theta_{\nu}{}^a - \Gamma^\rho_{\mu\nu}\, \theta_{\rho}{}^a + (\omega_{\mu})^{a}{}_{b}\, \theta_{\nu}{}^b = 0
\end{equation}
with $\Gamma^\rho_{\mu\nu}$, the Christoffel symbols $\left\{\substack{\rho\\ \mu\nu}\right\}$.  The two-form $\Omega^a{}_{b}$ parametrises the Riemann tensor
\begin{equation}
 \Omega^a{}_{b} = \textstyle{\frac{1}{2}} R^{a}{}_{bcd}\, \theta^c \wedge \theta^d.
\end{equation}

The main observation in the Sparling construction is that the vanishing of the Einstein tensor, i.e. the vacuum Einstein equation, is equivalent to the vanishing of
\begin{equation}
 E_{a} = \ast \Omega_{ab} \wedge \theta^b.
\end{equation}
Expanding out the expression above gives
\begin{equation}
 \ast \Omega_{ab} \wedge \theta^b = \frac{1}{4} \, \eta_{abcd} \, R^{cd}{}_{ef}\ \theta^b \wedge \theta^e \wedge \theta^f.
\end{equation}
Now, substituting the fact that
\begin{equation}
 \theta^b \wedge \theta^e \wedge \theta^f = \eta^{befg} \zeta_g
\end{equation}
for some one form $\zeta_a$ gives
\begin{equation}
 \ast \Omega_{ab} \wedge \theta^b = \ast R \ast_{ab}{}^{bc}\ \zeta_c.
\end{equation}
But, of course, $\ast R \ast_{ab}{}^{bc} = - G_a{}^c$, where $G_{ab}$ is the Einstein tensor contracted into the frame components.  In conclusion,
\begin{equation}
 E_a = - G_a{}^b \zeta_b.
\end{equation}
On the other hand, making use of the Cartan equations \eqref{cartan1} and \eqref{cartan2}, one can show that
\begin{equation} \label{sparling}
 E_a = \d W_a + S_a,
\end{equation}
where the two-form (or more generally $(d-2)$-form)
\begin{equation}
 W_a = \textstyle{\frac{1}{2}}\, \eta_{abcd} \ \omega^{bc} \wedge \theta^d
\end{equation}
is known as the Witten-Nester form.  It corresponds to the two-form integrated on the asymptotic boundary of a general spacelike hypersurface in Witten \cite{Witten} and Nester's \cite{Nester} proofs of the positive ADM mass theorem \cite{moreschi}.  The three-form (or more generally $(d-1)$-form) denoted $S_a$ is the Sparling form
\begin{equation}
 S_a = \textstyle{\frac{1}{2}} \eta_{abcd} \left( \omega^c{}_e\wedge \omega^{ed}\wedge \theta^b - \omega^{cd}\wedge\omega^b{}_e\wedge\theta^e \right).
\end{equation}
Note that while $E_a$ is clearly horizontal, $W_a$ and $S_a$ are not.  They depend on the particular choice of the orthonormal frame $\theta^a.$  From eqn.~\eqref{sparling}, we conclude that the Sparling form is exact if, and only if, the vacuum Einstein equation is satisfied.

\section{The linearised Einstein equation in the Sparling construction} \label{sec:pertspa}

In sections \ref{sec:ent} and \ref{sec:sparling}, we found that the Einstein equation (or its linearisation) can be related to the exterior derivative of a two-form in four dimensions and $(d-2)$-form in general.  In this section, we relate these two constructions by considering perturbations in the Sparling construction.  As in section \ref{sec:sparling}, we work in four dimensions.  However, our results will almost trivially generalise to higher dimensions. 

We consider a linearised perturbation on a background solution.  Accordingly, we split the vierbein into a background and perturbed piece
\begin{equation}
 \theta_\mu{}^a = e_{\mu}{}^a + f_{\mu}{}^a
\end{equation}
so that the perturbed part of the metric $h_{\mu\nu}$, defined via
\begin{equation}
 g_{\mu\nu} = \go_{\mu\nu} + h_{\mu\nu},
\end{equation}
is equal to
\begin{equation} \label{h}
 h_{\mu\nu} = 2\, e_{(\mu}{}^a f_{\nu) a}.
\end{equation}
Henceforth, all equations will be written to first order in the perturbation parameter.  The inverse vielbein
\begin{equation}
 \theta^\mu{}_a = e^\mu{}_a - e^\mu{}_b e^\nu{}_a f_{\nu}{}^b.
\end{equation}

Similarly, the spin connection decomposes as
\begin{equation}
 \omega^{a}{}_b = \omo{}^a{}_b + a^a{}_b,
\end{equation}
where the perturbed piece
\begin{align} \label{a}
 (a_\mu)^a{}_b = e^\nu{}_c e^\tau{}_b\, f_{\tau}{}^c \, \partial_\mu e_{\nu}{}^a - e^\nu{}_{b}\, \partial_\mu f_{\nu}{}^a 
 &+ \Gao{}^{\rho}_{\mu\nu}\, f_{\tau}{}^c\, (e_\rho{}^a e^\nu{}_c e^\tau{}_b + e^{\tau \, a} e^\nu{}_b e_{\rho\, c}) \notag \\[2mm]
  &+ \textstyle{\frac{1}{2}}\, e^{\sigma\, a} e^\nu{}_b (\partial_\mu h_{\sigma \nu} + \partial_\nu h_{\sigma \mu} - \partial_\sigma h_{\mu \nu})
\end{align}
has been calculated using the metric compatibility condition \eqref{metcom}.

The objects in the Sparling equation \eqref{sparling} are constructed from the vielbein $\theta^a$ and the spin connection $\omega^a{}_b$.  Hence, also we can decompose these into background and perturbed pieces
\begin{equation}
 E_a = \Eo_a + \delta E_a, \qquad W_a = \Wo_a + w_a, \qquad S_a = \So_a + s_a
\end{equation}
with
\begin{equation} \label{spaeqn:bgd}
 \Eo_a = - \Go_a{}^b\, \zeo_b, \quad \Wo_a = \textstyle{\frac{1}{2}}\, \eta_{abcd} \ \omo{}^{bc} \wedge e^d, 
 \quad \So_a = \textstyle{\frac{1}{2}} \eta_{abcd} \left( \omo{}^c{}_e\wedge \omo{}^{ed}\wedge e^b - \omo{}^{cd}\wedge\omo{}^b{}_e\wedge e^e \right)
\end{equation}
and
\begin{gather}
 \delta E_a = - \Go_a{}^b\, \delta\zeta_b - \delta G_a{}^b\, \zeo_b, \label{pert:E} \\[2mm]
 \quad w_a = \textstyle{\frac{1}{2}}\, \eta_{abcd} \left( \omo{}^{bc} \wedge f^d + a{}^{bc} \wedge e^d \right), \label{pert:w} \\[2mm]
  s_a = \textstyle{\frac{1}{2}}\, \eta_{abcd} \Big( a^c{}_e\wedge \omo{}^{ed}\wedge e^b + \omo{}^c{}_e\wedge a^{ed}\wedge e^b + \omo{}^c{}_e\wedge \omo{}^{ed}\wedge f^b \notag \\[1mm]
  \hspace{45mm} - a^{cd}\wedge\omo{}^b{}_e\wedge e^e - \omo{}^{cd}\wedge a^b{}_e\wedge e^e - \omo{}^{cd}\wedge\omo{}^b{}_e\wedge f^e \Big).
  \label{pert:s}
\end{gather}
This splits the Sparling equation into a background piece, which the background quantities will satisfy, and most importantly a perturbed piece that gives the linearised Einstein equation $\delta G_{ab}$, which appears in the expression for $\delta E_a$, in terms of the exterior derivative of $w_a$ and the perturbed Sparling form $s_a$
\begin{equation} \label{pert:spa}
 - \delta G_a{}^b\, \zeo_b = \Go_a{}^b\, \delta\zeta_b + \d w_a + s_a.
\end{equation}
For a traceless, transverse perturbation the linearised Einstein tensor is simply the Lichnerowitz operator on $h_{\mu\nu}$, which coincides with the background wave equation for the components of $h_{\mu\nu}$.

\subsection{Perturbations on a flat background} \label{sec:flat}

Before we consider the anti-de Sitter case, which will allow us to relate the Sparling construction to the linearised Einstein equation derived in Ref.\cite{ent3}, we consider first the simplest case of a perturbation on a flat background.  Recall that the Sparling construction depends on the choice of basis.  We choose to work with the simplest basis for which the vielbein, viewed as a matrix, is the identity
\begin{equation}
 \theta_\mu{}^a = \delta_\mu{}^a + f_\mu{}^a.
\end{equation}
In this basis the background spin connection vanishes
\begin{equation}
 \omo{}^{ab} = 0
\end{equation}
and, of course 
\begin{equation}
 \Go_{ab} = 0.
\end{equation}
Plugging these expressions into the definitions above gives that the Sparling form vanishes
\begin{equation}
 S_a = 0
\end{equation}
and
\begin{equation}
  \delta G_a{}^b\, \zeo_b = \d \left( - \textstyle{\frac{1}{2}}\, \eta_{abcd} \,  a{}^{bc} \wedge \delta^d \right).
\end{equation}
Hence, for perturbations on flat space we find that the linearised Einstein equation is given by the exterior derivative of 2-form $w_a$ as given above.

This result is related to the fact that in the weak field approximation that we are considering here, the energy-momentum tensor of the field $h_{\mu\nu}$ is second-order.  Thus, at the linearised level $h_{\mu\nu}$ does not contribute to the total energy \cite{trautman}.

\subsection{Perturbations on an AdS background} \label{sec:ads}

Moving on to the AdS case, as before, we proceed by choosing a background vierbein.  By inspecting the background metric, AdS space in Poincar\'e coordinates \eqref{met:ads}, we choose
\begin{equation}
 e_\mu{}^a = \frac{\ell}{z} \delta_\mu{}^a.
\end{equation}
Moreover, we choose to work in radial gauge in which $h_{\mu z} = 0$.  Hence, we have the freedom to set
\begin{equation}
 f_{z}{}^a = f_{\mu}{}^{\hat{z}} = 0.
\end{equation}
Moreover, we are free to set
\begin{equation}
 f_{t}{}^{\hat{x}} = f_{t}{}^{\hat{y}} = f_{x}{}^{\hat{y}} = 0.
\end{equation}
In this basis,
\begin{equation}
 \omo{}^{ab} = \begin{cases}
                -\frac{1}{\ell} e^{i}, \; \quad a=i,\ b=\hat{z} \\
                \frac{1}{\ell} e^i,  \qquad a=\hat{z},\ b=i \\
                0, \quad \qquad \text{otherwise}
               \end{cases},
\end{equation}
where $a=(\hat{z}, i)$.  Similarly, the only non-vanishing components of $\Gao{}^{c}_{ab}$ are
\begin{equation}
 \Gao{}^{a}_{az} = \Gao{}^{a}_{za} = -\frac{1}{z}, \quad \eta^{ii} \Gao{}^{z}_{ii} = \frac{1}{z},
\end{equation}
where we do not sum over repeated indices in the expressions above.  The background Einstein tensor
\begin{equation}
 \Go_{ab} = \frac{3}{\ell^2}\, \eta_{ab}.
\end{equation}
Now, we can go ahead and substitute all these quantities into the eqns.\ \eqref{spaeqn:bgd}--\eqref{pert:s} derived before.  However, the expressions we would obtain would not be as simple as those derived for the flat case in the previous section.  Therefore, we focus on the set-up considered in section \ref{sec:ent}.  We consider a hypersurface $\Sigma=\{t=t_0\}$ and investigate the $\hat{t}$-component of the pseudo-tensors that appear in the Sparling equation \eqref{sparling}.  This will allow us to derive the analogue of eqn.~\eqref{ent:Einstein} in the Sparling formalism.

Before we go on to look at the perturbed quantities, which includes the linearised Einstein equation, let us briefly verify that the background Sparling equation is indeed satisfied, as expected.  Working in conventions in which $\eta_{0123}=1$, where, henceforth we identify $\{\hat{t}, \hat{x}, \hat{y},\hat{z}\}$ with $\{0,1,2,3\}$, we find that on $\Sigma$
\begin{equation}
 \Eo_{0} = -\frac{3}{\ell^2}\, e^1 \wedge e^2 \wedge e^3, \qquad \Wo_{0} = \frac{2}{\ell}\, e^1 \wedge e^2, \qquad \So_0 = \frac{1}{\ell^2}\, e^1 \wedge e^2 \wedge e^3.
\end{equation}
Using the fact that
\begin{equation}
 \d \Wo_0 = -\frac{4}{\ell^2} \, e^1 \wedge e^2 \wedge e^3
\end{equation}
it is clear that
\begin{equation}
 \Eo_{0} = \d \Wo_0 + \So_0.
\end{equation}
Note, in particular, that $\So_{0}$ can be written as an exact form
\begin{equation}
 \So_0 = \d \left[ - \half \ell^{-1} e^1 \wedge e^2 \right].
\end{equation}

Next, let us consider the perturbed quantities. From eqn.~\eqref{pert:E},
\begin{equation}
 \delta E_0 = - \delta G_{00}\, \zeo{}^0 - \frac{3}{z^2} (f_x{}^1 + f_y{}^2)\, \d x \wedge \d y \wedge \d z.
\end{equation}
Similarly, from eqn.~\eqref{pert:w}
\begin{equation} \label{w0:1}
 w_0 = 3 a^{[12} \wedge e^{3]} + \frac{1}{z}\, (f_x{}^1 + f_y{}^2)\, \d x \wedge \d y.
\end{equation}
A straightforward calculation using the fact that
\begin{equation} \label{aeqns}
 (a_x)^{13} = \frac{z}{\ell}\, \partial_z f_x{}^1, \qquad (a_y)^{23} = \frac{z}{\ell}\, \partial_z f_y{}^2,
\end{equation}
gives that $s_0$ as defined in eqn.~\eqref{pert:s} reduces to
\begin{equation}
 s_0 = - \frac{1}{z}\, \partial_z (f_x{}^1 + f_y{}^2)\, \d x \wedge \d y \wedge \d z.
\end{equation}
Substituting the above expressions into eqn.~\eqref{pert:spa} with $a=0$ and simplifying gives
\begin{equation} \label{G00}
 - \delta G_{00}\, \zeo{}^0 = \d \left( 3 a^{[12} \wedge e^{3]} \right) + \frac{2}{z^2} (f_x{}^1 + f_y{}^2)\, \d x \wedge \d y \wedge \d z.
\end{equation}
From the definition of one-form $a^{ab}$ \eqref{a}, we find that
\begin{gather}
 (a_z)^{a3}=0,\qquad (a_x)^{12} = \frac{z}{\ell}\, (\partial_y f_x{}^1 - \partial_x f_y{}^1), \qquad (a_y)^{12} = - \frac{z}{\ell}\, \partial_x f_y{}^2.
\end{gather}
Together with equations \eqref{aeqns}, they give that eqn.~\eqref{G00} reduces to
\begin{align}
 - \delta G_{00}\, \zeo{}^0 = \d \Big[ (\partial_y f_x{}^1 - \partial_x f_y{}^1)\, \d x \wedge \d z - \partial_x f_y{}^2\, \d y \wedge \d z 
 - \partial_z (f_x{}^1 + f_y{}^2)\, \d x \wedge \d y \Big] \notag \\[2mm]
 + \frac{2}{z^2} (f_x{}^1 + f_y{}^2)\, \d x \wedge \d y \wedge \d z.
\end{align}
In fact, the above equation simplifies to
\begin{equation} \label{Gttspa}
  \delta G_{tt}\, \zeo{}^t = \d \left\{- \frac{\ell}{z} \left[ (\partial_y f_x{}^1 - \partial_x f_y{}^1)\, \d x \wedge \d z - \partial_x f_y{}^2\, \d y \wedge \d z 
 - \frac{\partial_z[z (f_x{}^1 + f_y{}^2)]}{z}\, \d x \wedge \d y \right] \right\}.
\end{equation}
Comparing the expression above for the linearised Einstein equation with that which appears in section \eqref{sec:ent}, eqn.~\eqref{ent:Einstein}, we identify $\zeo{}^t$ with $\epsilon^t$, the volume form on the hypersurface $\Sigma$: 
\begin{equation}
 \zeo{}^t = - \epsilon^t
\end{equation}
Moreover, we expect the two two-forms that appear on the right hand side of these respective equations to be related, possibly up to an exact one-form and this is what we show in the following.  

From the definition of $h_{\mu\nu}$ \eqref{h}, it follows that
\begin{equation}
 h^{\hat{\alpha}}{}_{\hat{\alpha}} = 2 \frac{z}{\ell}\, (f_x{}^1 + f_y{}^2).
\end{equation}
Moreover,
\begin{align}
 \partial^x h^{\hat{\alpha}}{}_{\hat{\alpha}} - \go{}^{xx} \partial^{\hat{\alpha}} h_{x \hat{\alpha}} &= \frac{z^3}{\ell^3} \left( 2 \partial_x f_y{}^2 - \partial_y f_y{}^1 \right), \notag \\[2mm]
 \partial^y h^{\hat{\alpha}}{}_{\hat{\alpha}} - \go{}^{yy} \partial^{\hat{\alpha}} h_{y \hat{\alpha}} &= \frac{z^3}{\ell^3} \left( 2 \partial_y f_x{}^1 - \partial_x f_y{}^1 \right).
\end{align}
and from eqn.~\eqref{eps2}
\begin{equation}
 \d x \wedge \d z = - \frac{z^4}{\ell^4} \epsilon_{ty}, \qquad \d y \wedge \d z = \frac{z^4}{\ell^4} \epsilon_{tx}, \qquad \d x \wedge \d y = \frac{z^4}{\ell^4} \epsilon_{tz}.
\end{equation}
Using the above equations, eqn.~\eqref{Gttspa} can be written as
\begin{align} \label{spa:Einstein}
 \delta G_{tt}\, \epsilon{}^t =\half \, \d \left\{   \partial_z h^{\hat{\alpha}}{}_{\hat{\alpha}}\, \epsilon^{t}{}_z - \left( \partial^{\hat{\alpha}} h^{\hat{\beta}}{}_{\hat{\beta}} -  \partial^{\hat{\beta}} h^{\hat{\alpha}}{}_{\hat{\beta}} \right) \epsilon_{t \hat{\alpha}} - \d \left[ \frac{\ell}{z} f_y{}^1 \d z \right] \right\}.
\end{align}
Now comparing the equation above, derived from the Sparling construction and eqn.~\eqref{ent:Einstein}, derived from the first law, we find that the two two-forms whose exterior derivatives gives the linearised Einstein equation match up to an exact term
\begin{equation}
 \d \left[ \frac{\ell}{z} f_y{}^1 \d z \right] = \d (h_{xy}\, \d z).
\end{equation}

\section{Conclusions}

We have found the potential in Ref.\cite{ent3} whose exterior derivative gives the linearised Einstein equation (or more precisely its $tt$-component) from the Sparling formalism.  Whereas in the case of perturbations on flat space, we found that the Sparling form vanishes and the exterior derivative of the Witten-Nester form gives the linearised Einstein equation, even for a background as simple as anti-de Sitter we were not able to make as general a statement and had to instead consider an ADM slicing of the spacetime.  In this case, the Sparling form on the ADM hypersurface becomes (off-shell) exact.  An obvious question is under what conditions the Sparling form vanishes or becomes off-shell exact?  Furthermore, can one gain a geometric understanding of why this happens?  An equivalent question is under what conditions (symmetry or otherwise) can the Einstein tensor, and hence the vacuum Einstein equation, be written as the exterior derivative of some $(d-2)$-form?

Going back to the main motivation of the paper emphasised in the introduction, i.e.\ the relationship between the holographic emergence of gravity on the one hand and suitable definitions of energy in general relativity on the other, there are hints of how one may proceed.  Focusing on asymptotically flat spaces, we have shown that from the Sparling construction perspective, the two-form that yields the linearised Einstein equation is the Witten-Nester two-form (see section \ref{sec:flat}). However, we know \cite{Witten, Nester, moreschi} that for an asymptotically flat space the integral of this two-form over asymptotic spacelike infinity gives the ADM mass.~\footnote{In fact, one may recognise that the second term in eqn.\ \eqref{spa:Einstein} multiplying the two-form $\epsilon_{t \hat{\alpha}}$ is precisely the same in structure as that which one would integrate to find the ADM mass.  The coincidence of these two expressions here is more than notational.}  On the other hand, from the Iyer-Wald formalism (see, in particular, equation \eqref{SEchi}) we know \cite{iyer} that the integral of this two-form over asymptotic spacelike infinity gives the canonical energy, which coincides with the ADM mass \cite{iyer}.  Thus, we identify the Iyer-Wald two-form with the Witten-Nester two-form.  Beyond the scope of asymptotically flat spaces, we have demonstrated in this paper that the same correspondence holds for asymptotically AdS spaces.  A possible application of these ideas could be in the context of flat space holography.  

Within the context of AdS holography, can the relation with the Sparling formalism, which gives the full non-linear Einstein tensor, allow one to do better than to derive simply the linearised Einstein equation from holographic arguments?  In many respects, the full, non-linear, Sparling construction is much simpler and intuitive than the linearised version, which we derived here.  This fact gives rise to reasonable optimism that the Sparling formalism has much more to say about holography and the emergence of bulk locality.

\section*{Acknowledgements}
I am indebted to David Skinner for discussions and remarks that initiated this study.  This work is partially supported by grant no. 615203 from the European Research Council under the FP7.

\bibliographystyle{utphys}
\bibliography{sparling}

\end{document}